# Specific heat and magnetocaloric effect in $Pr_{1-x}Ag_xMnO_3$ manganites

A. G. Gamzatov*, A. M. Aliev**, A. B. Batdalov

*Institute of Physics, Dagestan Scientific Center, Russian Academy of Sciences, Makhachkala, 367003 Russia*
**e-mail: gamsatov_adler@mail.ru*
***e-mail: lowtemp@mail.ru*

H. Ahmadvand, H. Salamati, P. Kameli

*Department of Physics, Isfahan University of Technology, 84156-83111 Isfahan, Islamic Republic of Iran*

**Abstract**

The magnetocaloric effect in alternating magnetic fields has been investigated in $Pr_{1-x}Ag_xMnO_3$ manganites with $x$=0.05-0.25. The stepwise reversal of the sign of the magnetocaloric effect has been revealed in a weakly doped sample ($x$=0.05) at low temperatures (~80 K). This reversal is attributed to the coexistence of the ferromagnetic and canted antiferromagnetic phases with different critical temperatures.

The physical properties of $Pr_{1-x}Ag_xMnO_3$ praseodymium manganites (A is a univalent metal Na, K, Ag, etc.) are strongly different from the properties of manganites of other rare_earth metals (La, etc.). This system does not undergo a metal–insulator phase transition characteristic of other manganites and its existence region is much narrower, $x \leq 0.25$ [1–3]. Moreover, the Curie temperature $T_C$ of this system is lower than 136 K [2] owing to a large difference between the ion radii of Pr and substituting cations Na, K, and Ag ($r^{Pr}$ =0.099 nm, $r^{Na}$=0.116 nm, $r^{K}$=0.138 nm, and $r^{Ag}$=0.115 nm), which causes local distortions of the crystal lattice; a decrease in the Mn–O–Mn valence angle; a weakening of the exchange interaction; and, as a result, a decrease in $T_C$ [4].

The effect of the substitution of univalent alkali metal atoms Na and K for Pr atoms on the electric and magnetic properties was analyzed in [1, 2]. The structure and magnetic and electric properties of $Pr_{1-x}Ag_xMnO_3$ were investigated in [3]. According to [3], the behavior of the electric resistivity $\rho(T)$ down to low temperatures is of a semiconducting character and the temperature dependence of the magnetic susceptibility has anomalies characteristic of the paramagnet–ferromagnet phase transition. For this reason, $Pr_{1-x}Ag_xMnO_3$ was classified as a ferromagnetic insulator whose Curie temperature depends on the doping level.

In this work, we report the experimental results for the specific heat $C_P$ and magnetocaloric effect $\Delta T$ of the $Pr_{1-x}Ag_xMnO_3$ ceramic samples with $x$=0.05, 0.1, 0.15, and 0.25 for temperatures $T$=77–300 K. The technology of the manufacture of the samples and their magnetic and electric properties were described in [3]. According to the measurements of the magnetic susceptibility, the paramagnet–ferromagnet phase transition is observed in all of the samples and $T_C$ depends nonmonotonically on the doping level: it increases sharply at weak doping, reaches a maximum at $x$=0.15–0.20, and decreases smoothly with a further increase in the doping level.

The specific heat was measured by the ac calorimetry method [5] and, to directly measure the magnetocaloric effect, a special method based on the measurement of the amplitude of sample temperature oscillations in the presence of a weak ac magnetic field was developed [6].

The essence of this method is as follows. The external alternating magnetic field $H=H_0\cos\omega t$ ($H_0$ is the amplitude and $\omega$ is the cyclic frequency) applied to a magnetic material induces oscillations of its temperature $T=T_0\cos(\omega t+\varphi)$, where $\varphi$ is the phase shift of oscillations of the temperature with respect to oscillations of the magnetic field. These oscillations are detected by a chromel–constantan thermocouple glued to a sample. To improve the thermal contact of the thermocouple with the sample and reduce inertia, the thermojunction is compressed to 3–5 μm. An ac signal from the thermocouple is detected with a high accuracy by a phase-sensitive nanovoltmeter. In this work, the measurements were performed at frequencies of 0.3 to 0.5 Hz. The alternating magnetic field with amplitudes from 0 to 1 kOe was generated by means of an electromagnet and a power supply unit with external control. The control ac voltage was fed to the power supply unit from the phase-sensitive SR830 nanovoltmeter. This procedure makes it possible to detect temperature changes with an accuracy of no worse than 10–3 K. For calibration, the magnetocaloric effect of a Gd single-crystal sample was measured.

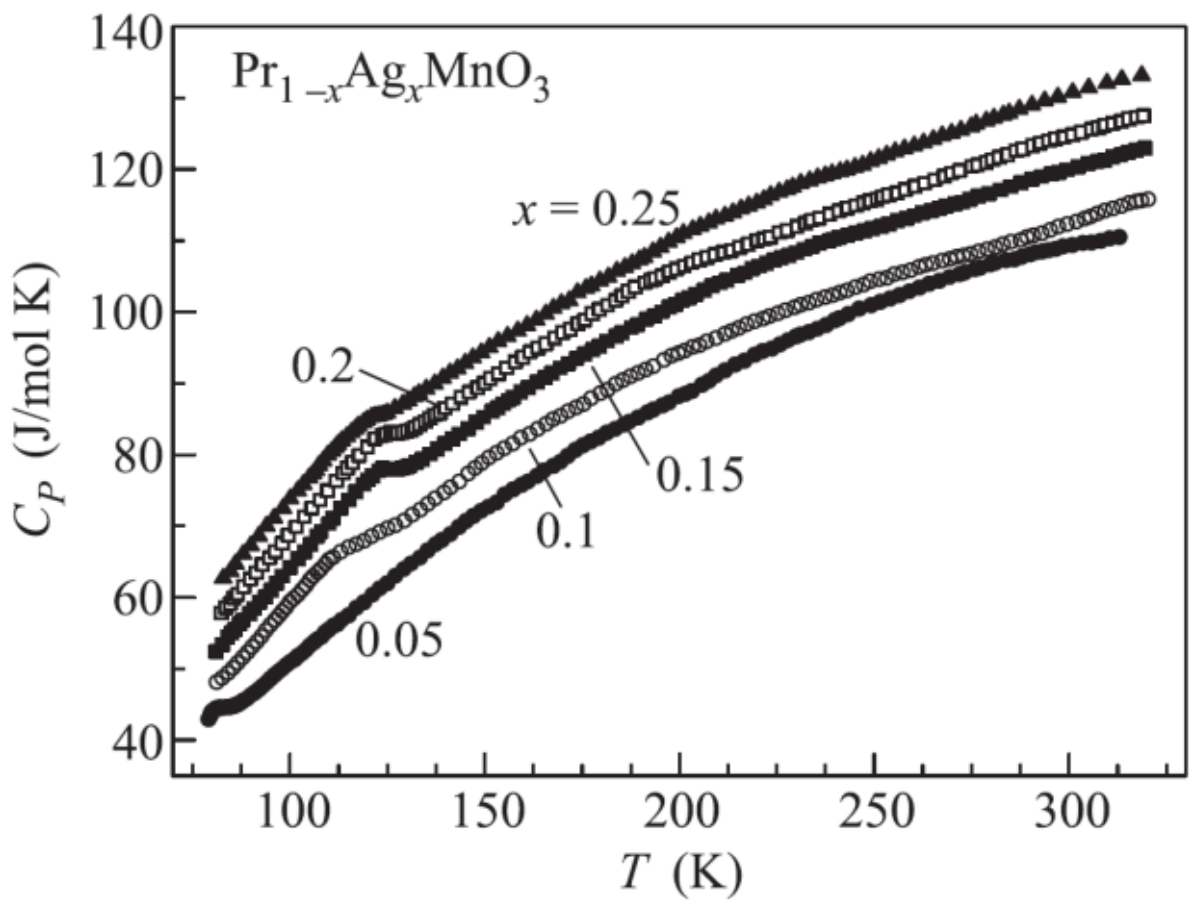

**Fig. 1.** Temperature dependence of the specific heat of the $Pr_{1-x}Ag_xMnO_3$ samples with $x$=0.05, 0.1, 0.15, 0.2, and 0.25. For clarity, the curves for $x$≤0.05 are shifted relative to the curve for $x$=0.05.

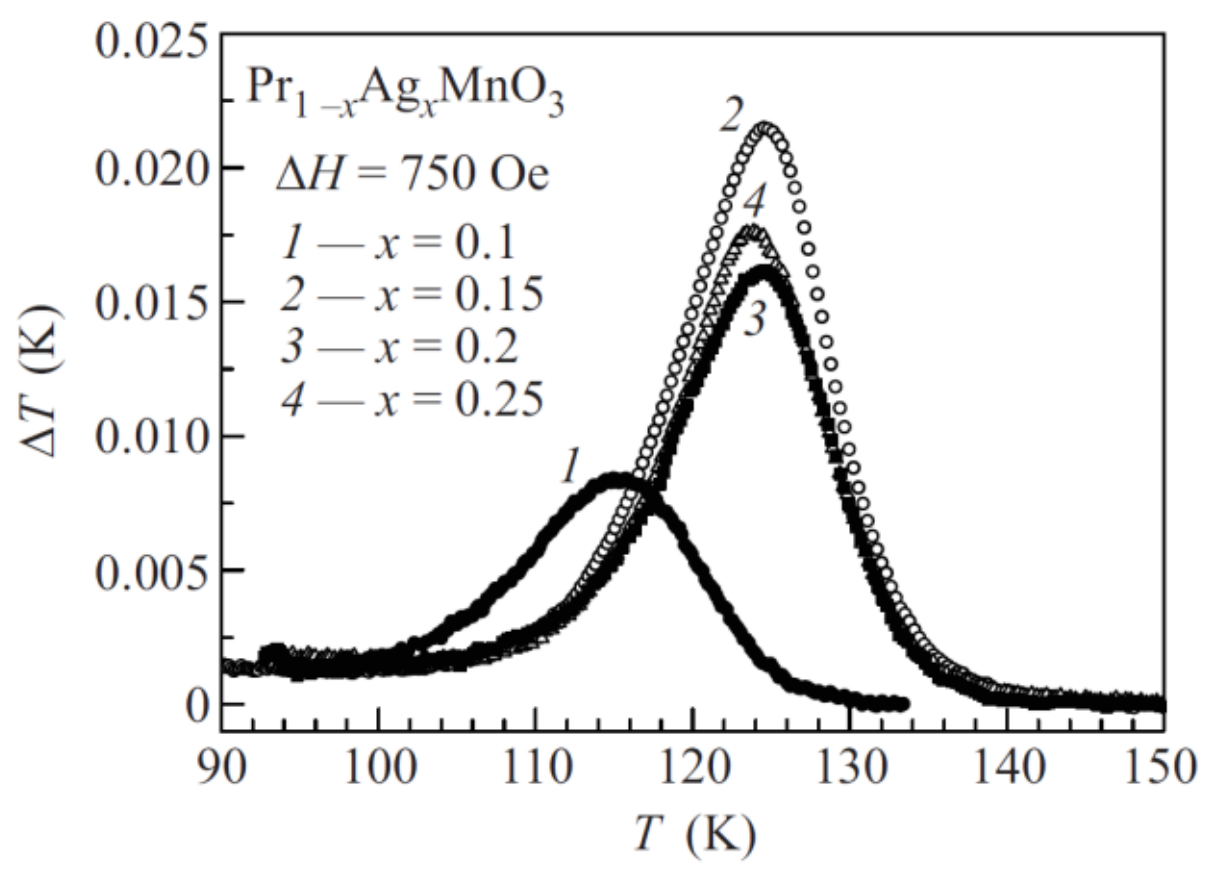

**Fig. 2.** Temperature dependence of the magnetocaloric effect $\Delta T$.

Figure 1 shows the temperature dependence of the specific heat of the samples under investigation. It is seen that the region of the magnetic phase transition is manifested as small anomalies, which indicate the smearing of the phase transition and the possible chemical inhomogeneity of the samples. Small anomalies of the specific heat show that these samples have low transition entropies. This means that only a small part of the sample transits to a magnetically ordered state.

Figure 2 shows the temperature dependence of the magnetocaloric effect of the samples under investigation. As is seen, for $x$≥0.1, the observed magnetocaloric effect is typical of ferromagnets: the effect increases with a decrease in the temperature, reaches a maximum near $T_C$, and decreases smoothly with a further decrease in the temperature. The maximum effect is observed for $x$=0.15, which corresponds to the optimal doping level for manganites doped with univalent metals.

The temperatures of the phase transitions estimated from the data for the magnetocaloric effect and specific heat are lower than the values obtained from the measurements of the magnetic susceptibility [3], but correlate with them (see Fig. 3).

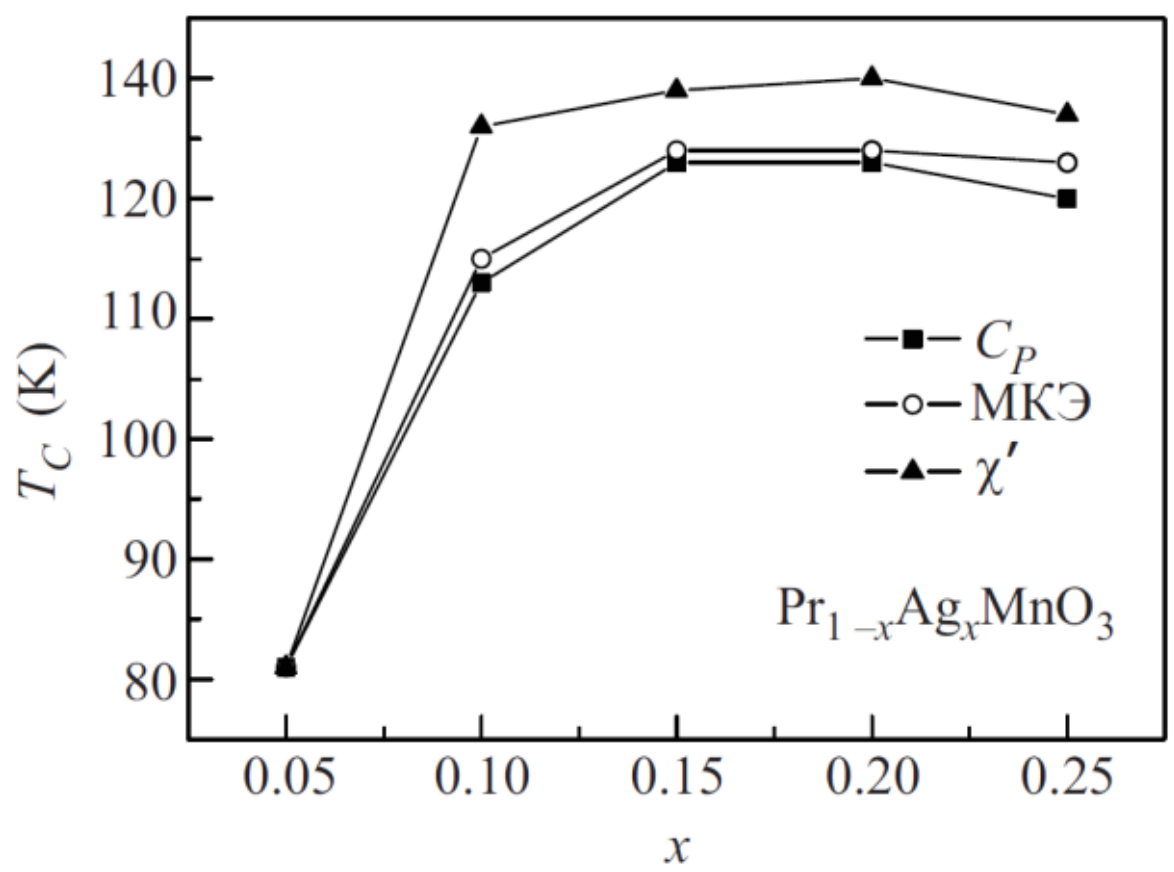

**Fig. 3.** Curie temperature $T_C$ versus the doping level $x$ according to the data on the (squares) specific heat, (circles) magnetocaloric effect, and (triangles) susceptibility.

The maximum $\Delta T$ value recalculated per unit magnetic field intensity under the assumption of the linear dependence of $\Delta T$ on $\Delta H$ is 0.029 K/kOe, which is much smaller than similar values for other univalent substituted manganites (0.104 [7] and 0.085 K/kOe [8] for $La_{1-x}Ag_xMnO_3$ and 0.21 K/kOe [9] for $La_{1-x}K_xMnO_3$).

The behavior of the magnetocaloric effect in the $Pr_{0.95}Ag_{0.05}MnO_3$ sample (see Fig. 4) is particularly remarkable. As is seen in the figure, the direct magnetocaloric effect characteristic of ferromagnets is first observed with a decrease in the temperature; it reaches a maximum at about $T_m$=82.3 K. In a narrow temperature range $\Delta T \approx 0.4$ K near $T \approx 80$ K, the stepwise reversal of the sing of the magnetocaloric effect occurs. With a further decrease in the temperature, the absolute value of the negative magnetocaloric effect reaches a maximum at $T_m$=78.7 K and, then, decreases sharply.

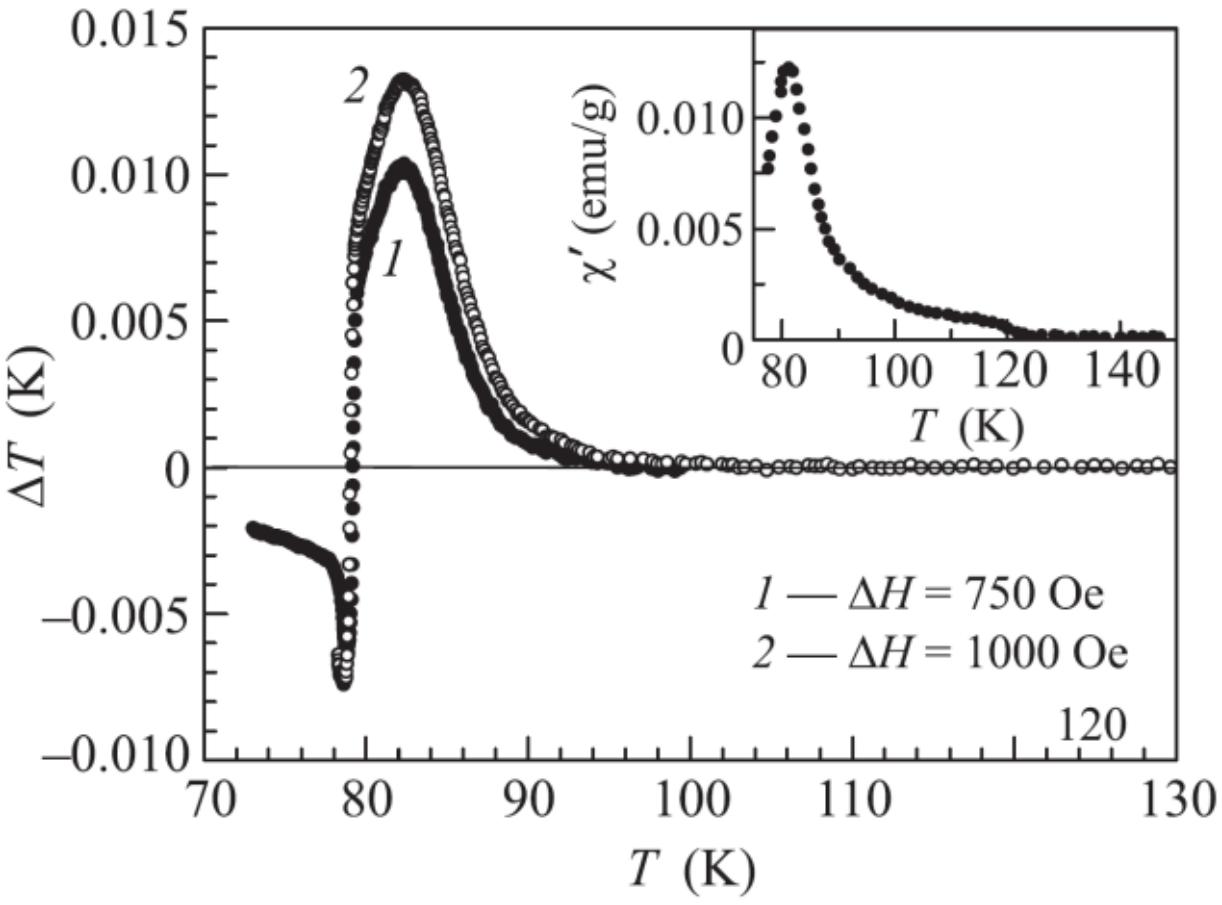

**Fig. 4.** Temperature dependence of the magnetocaloric effect in the $Pr_{0.95}Ag_{0.05}MnO_3$ sample. The inset shows the temperature dependence of the susceptibility of this sample [3].

The reversal of the sign of the magnetocaloric effect from positive to negative means that the antiferromagnetic interaction begins to dominate below this temperature. These data are in agreement with the data for the magnetic susceptibility [3] (see inset in Fig. 4). The temperature dependence of the susceptibility has two anomalies at 119 and 81 K corresponding to the transition to the ferromagnetic (at $T_C$) and canted antiferromagnetic states, respectively. The approximation of the magnetic susceptibility above the temperatures of the maxima $T_{can}$ and $T_C$ by the Curie–Weiss law $\chi = C/(T-\Theta)$ provides the positive constants $\Theta_1$ = 118 K and $\Theta_2$ = 82 K, which indicate the ferromagnetic nature of the interaction in both transitions. In view of this circumstance, the reversal of the sign of the magnetocaloric effect is a surprising effect. It is worth noting that a small anomaly on the $\chi(T)$ curve observed near $T$=119 K in [3] is not manifested in our experiments in spite of the high accuracy of the measurements of $\Delta T$ and $C_P$. This anomaly is assumingly a feature of a particular sample and is not inherent in the system on the whole.

At the same time, with an increase in the doping level of the pure PrMnO3 compound with univalent ions, the type of magnetic interaction changes from antiferromagnetic to ferromagnetic and this transition is discontinuous. According to the data reported in [1], the canted antiferromagnetic phase is characteristic of the $Pr_{1-x}Na_xMnO_3$ samples with $x$=0.025 and 0.05, whereas the canted antiferromagnetic phase $A_yF_z$ and the pure ferromagnetic phase $F_y$ with the critical temperatures $T_C$=106 K and $T_N$=80 K, respectively, coexist in the sample with $x$=0.075. The pure ferromagnetic transition is observed for $x$ >0.1 as in our case.

The observed temperature dependence of the magnetocaloric effect is adequately explained under the assumption that both phases with different, but close critical temperatures ($T_C > T_N$) coexist in the $Pr_{0.95}Ag_{0.05}MnO_3$ sample: the magnetocaloric effect is positive ($\Delta T > 0$) and negative ($\Delta T < 0$) near the ferromagnetic transition ($T_C$=82.7 K) and $T_N$=78.7 K, respectively.

In other words, ferromagnetic and antiferromagnetic phases with very close transition temperatures coexist in the weakly doped $Pr_{1-x}Ag_xMnO_3$ system in a limited temperature range. Their ratio varies with the temperature and the measured magnetocaloric effect is the difference between the antiferromagnetic and ferromagnetic contributions to the magnetocaloric effect. The possibility of the appearance of such a state in manganites was pointed out in [10].

To conclude, the reversal of the sign of the magnetocaloric effect has been revealed in $Pr_{1-x}Ag_xMnO_3$ manganites. This reversal can be explained by the coexistence of the ferromagnetic and canted antiferromagnetic phases whose volume fractions vary with the temperature. The critical temperatures estimated from the data for the specific heat and magnetocaloric effect correlate with the published data obtained from the magnetic measurements.

This work was supported by the Russian Foundation for Basic Research (project no. 09-08-96533) and by the Branch of Physical

Sciences, Russian Academy of Science (program “Strongly Correlated Electrons in Solids and Structures”).